\documentclass[epj-spec]{svjour}
\usepackage{graphicx}
\usepackage{hyphenat}
\begin{document}
\bibliographystyle{epj}
\title{Muon capture on the proton and deuteron}
\author{Frederick Gray\inst{1}\fnmsep\thanks{\email{fgray@regis.edu}} 
for the MuCap~\cite{muCapCollaboration} and MuSun~\cite{muSunCollaboration}
Collaborations}
\institute{\inst{1}Department of Physics, Regis University, 3333 Regis Blvd., Denver, CO~~80221, U.S.A.}
\abstract{
By measuring the lifetime of the negative muon in pure protium ($^1$H),
the MuCap experiment determines the rate of muon capture on the proton,
from which the proton's pseudoscalar coupling $g_p$ may be inferred.
A precision of 15\% for $g_p$ has been published; this is a step along the way 
to a goal of 7\%.  This coupling can be calculated precisely from heavy baryon 
chiral perturbation theory and therefore permits a test of QCD's chiral 
symmetry.  Meanwhile, the MuSun experiment is in its final design stage; it 
will measure the rate of muon capture on the deuteron using a similar 
technique.  This process can be related through pionless effective field 
theory and chiral perturbation theory to other two-nucleon reactions of 
astrophysical interest, including proton-proton fusion and deuteron breakup.
} %end of abstract
\maketitle
\section{Introduction}
\label{intro}

Muon capture on the proton,
\begin{equation}
\label{eq:mup}
\mu^- + p \rightarrow \nu_\mu + n ~,
\end{equation}
is a fundamental hadronic weak-interaction process.  By measuring the
rate at which it occurs, one can constrain the rich structure of the proton 
that arises from the interactions of quarks and gluons within it. 
On the other hand, muon capture on the deuteron,
\begin{equation}
\label{eq:mud}
\mu^- + d \rightarrow \nu_\mu + n + n~,
\end{equation}
is among the simplest two-nucleon processes, and can therefore contribute to 
the understanding of similar interactions.  Notably, it may be connected 
to the fusion reaction $p + p \rightarrow d + e^+ + \nu_e$ that fuels 
stars (including our sun) and to the deuteron breakup reactions used by 
the Sudbury Neutrino Observatory to study neutrino oscillations.

Theoretical calculations that reflect the chiral symmetry of quantum
chromodynamics (QCD) have been developed to describe 
processes~\ref{eq:mup} and~\ref{eq:mud}.  These calculations have reached
impressive levels of precision, and they await experimental input that 
the MuCap and MuSun collaborations will soon provide.

\section{Muon capture on the proton: theoretical motivation}

Muon capture arises from the interaction between a
leptonic current (representing the muon's transformation into its neutrino) 
and a hadronic current (the proton's transformation into a neutron).
The hadronic current may be parameterized by writing down terms with
all possible 

\noindent Lorentz-invariant symmetry properties:
\begin{eqnarray}
\label{eq:hadcurrent}
J_\alpha & = & g_v(q^2)\gamma_\alpha - g_a(q^2)\gamma_\alpha\gamma_5 
              + g_m(q^2) \frac{i}{2 m_N} \sigma_{\alpha\beta} q^\beta \nonumber \\
   & &        -~g_t(q^2) \frac{i}{2 m_N} \sigma_{\alpha\beta} q^\beta \gamma_5
              + g_s(q^2) \frac{1}{m_\mu} q_\alpha - g_p(q^2) \frac{1}{m_\mu} q_\alpha \gamma_5 ~.
\end{eqnarray}
In this expression, the terms correspond to vector, axial vector,
weak magnetic, tensor, scalar, and pseudoscalar components.
Our interest is in the pseudoscalar part.  Fortunately, the others
are either well-determined experimentally ($g_v$, $g_a$, and $g_m$) 
or can be shown theoretically to be zero by a $G$-parity symmetry 
argument (the ``second class'' currents $g_t$ and $g_s$).  Therefore, a 
measurement of the rate of muon capture at rest in protium is effectively a 
measurement of $g_p$ at the fixed $q^2 = -0.88 m_\mu^2$ that corresponds to 
the kinematics of the process.

An effective field theory (EFT) approach to low-energy phenomena 
parameterizes the higher-energy physics that has been abstracted 
from the theory as a set of low-energy constants.  It then proceeds
with a series expansion in a small parameter $Q/\Lambda$ to describe
phenomenological observables.  With zero quark masses, the QCD Lagrangian 
is chirally symmetric: that is, the left- and 
right-handed components of the Dirac spinor are treated identically.  This 
symmetry is spontaneously broken by the addition of nonzero quark mass terms.  
Heavy baryon chiral perturbation theory, an EFT that expands in both 
the pion (or light quark) mass and the coupling constant, takes advantage
of this near-symmetry of the QCD.  It leads to a theoretical 
prediction~\cite{Bernard:1995dp} for $g_p$:
\begin{equation}
\label{eq:gp}
g_p(q^2) = \frac{2 m_\mu g_{\pi N N}(q^2) F_\pi}{m_\pi^2 - q^2} - \frac{1}{3} g_A(0) m_\mu m_N r_A^2
\end{equation}
which becomes $8.26~\pm~0.23$ when evaluated at the characteristic $q^2$
for muon capture.  This result agrees with earlier results based on 
the partially conserved axial current (PCAC) and current 
algebra~\cite{Gorringe:2002xx}.
Thus, a fundamental symmetry of the Standard Model predicts $g_p$ with
a precision of 3\%.

\section{Muon capture on the proton: experimental status}

Meanwhile, the precise theoretical prediction for $g_p$ has long eluded an 
equally precise experimental check.  The final state of muon capture 
contains no charged particles, so it is difficult to reliably detect directly.
Experiments using neutron detection for muons stopped in a hydrogen
gas target (performed in the late 1960s and early 1970s) achieved precisions 
of 9\%~\cite{AlberigiQuaranta:1969ip} and 13\%~\cite{Bystritsky:1974}
for the capture rate.

\begin{figure}
\begin{center}
\includegraphics[width=0.75\columnwidth]{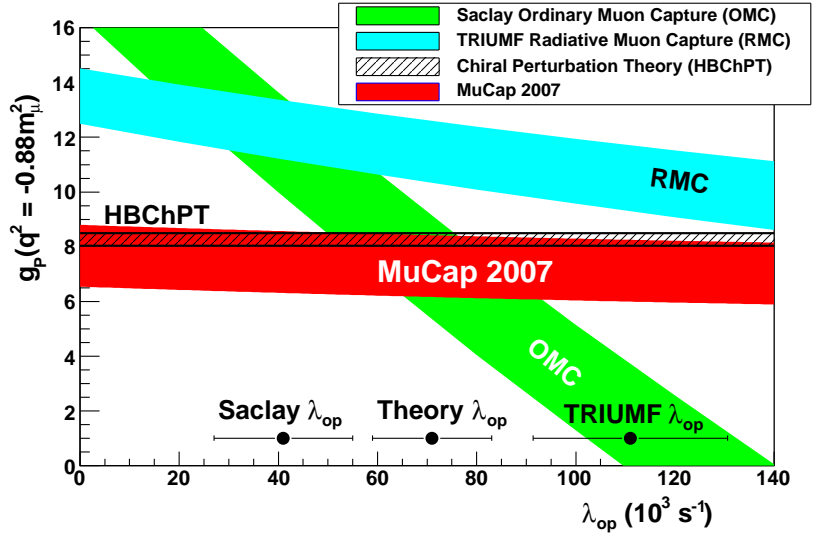} 
\end{center}
\caption{The proton's pseudoscalar 
coupling $g_p$ inferred from the observed capture rate in the
previous generation of experiments (green and cyan bands) depends strongly on 
the ortho-para transition rate $\lambda_{OP}$.  The MuCap experiment (red
band) reduces this dependence substantially by using a low-density gas target,
thereby confirming the heavy baryon chiral perturbation theory 
prediction (hatched band).}
\label{fig:gp_lop}       % Give a unique label
\end{figure}

Several more recent experiments have measured the muon capture rate 
by other techniques.  The Saclay ordinary muon capture
 measurement~\cite{Bardin:1980mi} compared the apparent lifetime 
$\tau_{\mu^-}$ of the negative muon to the lifetime $\tau_{\mu^+}$ of 
the positive muon.  The negative muon can 
decay ($\mu^- \rightarrow e^- + \bar\nu_e + \nu_\mu$) or it can be
captured, whereas positive muons can only decay. 
Attributing the additional 
part of the disappearance rate to the capture process, one obtains
for the capture rate 
\begin{equation}
\label{eq:disappearance}
\Lambda_S = \lambda_{\mu^-} - \lambda_{\mu^+} = \frac{1}{\tau_{\mu^-}} - \frac{1}{\tau_{\mu^+}} ~.
\end{equation}
The TRIUMF radiative muon capture measurement~\cite{Wright:1998gi} counted
the photons emitted in the rare process 
$\mu^- + p \rightarrow \nu_\mu + n + \gamma$.
However, both of these experiments were problematic because of the high 
density of the liquid hydrogen targets that they employed.  
In such targets,
the atomic and molecular kinetics of the muon-proton system become 
very important.  After a negative muon slows down through multiple scattering,
it becomes electromagnetically bound to a proton, forming a muonic hydrogen 
atom.  Such an atom behaves in many respects like ordinary hydrogen, but 
with a binding radius that is reduced by $m_\mu/m_e \approx$~207.
Initially, the singlet and triplet atomic hyperfine states are statistically
filled, but the triplet state is very quickly depopulated through
collisions with hydrogen molecules.  It is from the resulting pure atomic 
singlet state that we intend to measure the nuclear muon capture 
rate $\Lambda_S$.  However, at a rate 
of $\lambda_{of} \approx 2 \times 10^6$~s$^{-1}$, muonic hydrogen molecules 
($p-\mu-p$) are formed.  Initially, nearly all are in the 
ortho-molecular~($J=1$) state; over time, the population shifts 
to the para-molecular~($J=0$)
state at a rate given by $\lambda_{OP}$.  This parameter is very poorly-known;
independent measurements~\cite{Bardin:1981cq,Clark:2005as} (shown 
in Figure~\ref{fig:gp_lop}) have given substantially inconsistent results, 
and neither experiment agrees with the theoretical 
calculation~\cite{Bakalov:1980fm}.  The large uncertainty associated with 
$\lambda_{OP}$ makes it difficult to extract $g_p$ from the results of
either the Saclay or the TRIUMF experiment.

\begin{figure}
\begin{center}
\includegraphics[width=0.75\columnwidth]{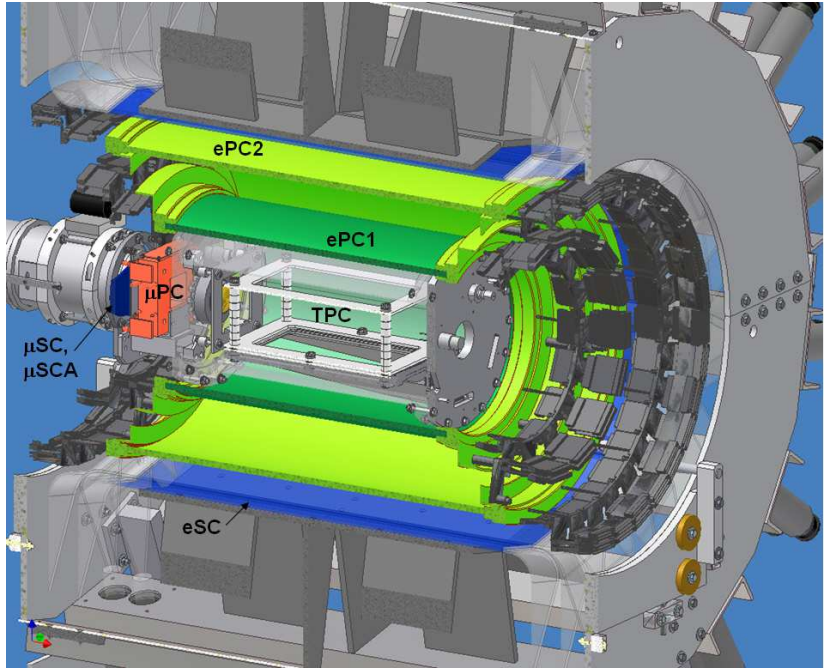}
\end{center}
\caption{The MuCap apparatus stops muons in a 10~bar protium environment, 
where the fiducial volume is surrounded by a time-projection chamber (TPC)
that determines the muon stopping point.  Electrons from muon decay within
the hydrogen are detected by the surrounding cylindrical layers of 
proportional chambers (ePC1 and ePC2) and plastic scintillator (eSC).}
\label{detscheme}       % Give a unique label
\end{figure}

The MuCap experiment, performed using a muon beam from the high-intensity
590~MeV proton cyclotron at the Paul Scherrer Institute, uses a protium
(isotopically pure $^1$H) gas target
whose density is $\phi=$1\% times that of liquid hydrogen.  At room
temperature, this density corresponds to 10~bar pressure.  Both the 
molecular formation rate and the ortho-para transition rate are 
proportional to $\phi$, so the dependence on the unknown $\lambda_{OP}$
parameter is dramatically suppressed.  In MuCap, 96\% of the
capture events come from the desired atomic singlet state.
The basic experimental technique 
is a disappearance rate measurement similar to that used in the Saclay 
experiment, described by Equation~\ref{eq:disappearance}.  Under these
conditions, the probability that a muon will be captured is 
approximately 0.16\%.  However, the low
density introduces a new complication: a significant number of the incident
muons actually stop outside the target gas, so an essential aspect of the 
experiment is to track each muon and include only those that can be proven 
to have stopped in hydrogen.  For this reason, the target gas 
constitutes the active medium of a time projection chamber (TPC) that
records the muon's ionization tracks.  The deposited charge drifts vertically
at a speed of 5.5~mm/$\mu$s in an applied electric field of 2~kV/cm, so the 
vertical coordinate of the track may be determined by its arrival time at the 
readout grid at the bottom.  The sensitive volume of the TPC 
measures $15\times12\times28$~cm$^3$; a Bragg peak is clearly visible
at the end of the track, and its three-dimensional coordinates are required
to fall at least 1.5~cm from any boundary.  As shown in Figure~\ref{detscheme},
the TPC is surrounded by a set of detectors (proportional chambers 
and scintillators) that track the electrons from muon decay.  
The disappearance rate is determined by a $\chi^2$ minimization of an 
exponential function (plus a constant background term) relative to the
spectrum of muon decay times.

The MuCap collaboration recently reported a first 
measurement~\cite{Andreev:2007wg} of the muon disappearance rate 
$\lambda_{\mu^-} = 455\,887.2 \pm 16.8$~s$^{-1}$.
To extract $\Lambda_S$, the positive muon lifetime is taken from the 
MuLan experiment, which performed by a partially overlapping collaboration and 
using the same beamline at PSI as MuCap.  An 11 part-per-million (ppm) result
for the lifetime of $\tau_{\mu+} = 2.197\,013(21)(11)~\mu$s has been 
published~\cite{Chitwood:2007pa}, with the intention to reach 1~ppm based 
on data already collected.  The value agrees well with a measurement 
made within the MuCap apparatus as a consistency check, which gave
$\tau_{\mu+} = 2.197\,01(14)~\mu$s.
When these results are combined, the ensuing
value for the capture rate is $725.0 \pm 17.4$~s$^-1$.  
From this value, $g_p(q^2=-0.88 m_\mu^2) = 7.3 \pm 1.1$ may be computed, 
once newly-computed radiative corrections~\cite{Czarnecki:2007th} are taken 
into account.  Within this 15\% precision, the prediction of HB$\chi$PT is 
confirmed, thereby resolving a longstanding puzzle of nuclear physics.

The purity of the hydrogen target gas is critical for this measurement;
since muon capture rates tend to scale as $Z^4$, a small impurity concentration
can lead to a large change in the total disappearance rate.
The experiment is fabricated from intrinsically clean,
low-outgassing materials; notably, the frame of the TPC is made from
Borofloat glass so that the chamber can be baked to 115$^\circ$~C without damage.
The gas is initially filled into the system through a palladium-foil filter.
It is then continuously circulated through a cooled Zeolite absorber 
system~\cite{Ganzha:2007uk} that was demonstrated to reduce the 
concentrations of N$_2$ and O$_2$ to less than 7 parts per billion (ppb) and 
the concentration of H$_2$O vapor to 18-30~ppb.  During the experiment,
the yield of muon capture on impurity nuclei was monitored in real time by 
the unique signature of recoiling nuclei in the TPC data.  Impurity-doped runs 
were collected in order to confirm the effect on the disappearance rate, and 
an appropriate correction was applied.

The protium gas used in MuCap must also be substantially depleted of deuterium.
The scattering cross section for a muon-deuteron atom in protium gas
has a Ramsauer-Townsend minimum at 1.6~eV, so any muon-deuteron
atoms that might be formed tend to diffuse long distances through the 
target gas.  This process represents a time-dependent loss of muons from 
the fiducial volume of the TPC, which in turn distorts the muon disappearance
spectrum.  Accelerator mass spectroscopy (AMS)
measurements~\cite{2007NIMPB.259....7S} have confirmed that the hydrogen used 
in the published data set, which was produced by electrolysis of 
deuterium-depleted water purchased from Ontario Power Generation, had a 
deuterium content of $1.44\pm0.13$~ppm.  This level agrees well with the 
dependence of the muon disappearance rate on the observed impact parameter
between the muon stopping point and the electron track.  Again, dedicated
runs with a deuterium-doped target allowed an appropriate correction 
to be made.  

The published MuCap result is based on the analysis of $1.6\times10^9$ 
stopped muon decays observed in the summer 2004 running period. 
This represents less than 10\% of the full MuCap data set, which is
anticipated to give a final precision of 7\% for $g_p$.  The rate at
which muons could be stopped was substantially enhanced by the
MuLan kicker~\cite{2004ITPS...32.1932B}, which allows 
``Muon on Demand'' operation.  For 25~$\mu$s after an entering muon has
been detected, the beamline is closed off by the application of a 
$\sim$1.6~kV/cm electric field along a 1.5~m segment upstream from the 
detector.   This time structure allows efficient operation by eliminating
``pile-up'' of muons within the TPC; each muon typically arrives just as
the ionization charge deposited by its predecessor has been cleared from the 
chamber.  The other substantial improvement in the experiment was to the 
purity of the target gas: a cryogenic isotope separation was performed, 
leading to a deuterium concentration less than 0.006~ppm (a conservative upper 
limit based on AMS results) in the hydrogen that was used 
for the forthcoming data.

\section{Muon capture on the deuteron}

Muon capture on the deuteron is a member of a family of interesting
reactions; other members include proton-proton 
fusion ($p + p \rightarrow d + e^+ + \nu_e$) and the charged and neutral 
current deuteron breakup reactions ($\nu_x + d \rightarrow p + n + \nu_x$ 
and $\nu_e + d \rightarrow n + n + e^-$).  Each of these is a semileptonic 
two-nucleon interaction that either forms or destroys a deuteron.  
The fusion process drives energy production in stars, while the deuteron 
breakup processes are used by the Sudbury Neutrino Observatory to monitor the 
flux of solar neutrinos~\cite{Aharmim:2007nv}.  Consequently, an
improved understanding of the cross-sections for these reactions will have
substantial astrophysical implications.
A new proposal~\cite{muSunProposal} to measure the rate of muon capture on 
the deuteron has been submitted to PSI and has received approval; 
the experiment will be known as MuSun.

The connections among these processes are often discussed in terms of 
a {\em pionless} EFT, which treats the pion as a high-energy particle to 
be integrated into the low-energy constants (LECs).  The set of LECs consists 
of many that are well-known from one-nucleon processes, plus a new 
constant $L_{1A}$ that describes the two-nucleon weak axial current.   
The determination of $L_{1A}$ with by far the smallest stated uncertainty 
is $4.2\pm0.1$~fm$^3$ from the rate of tritium beta 
decay (\cite{Schiavilla:1998je} as interpreted by~\cite{Chen:2002pv}).  
However, because the triton is a three-nucleon 
system, this calculation may introduce a significant model dependence.
Other relevant techniques include reactor $\bar\nu + d$ scattering 
experiments~\cite{Butler:2002cw} ($L_{1A} = 3.6\pm5.5$~fm$^3$), 
self-consistency of the solar neutrino 
data~\cite{Chen:2002pv} ($L_{1A} = 4.0\pm6.3$~fm$^3$), and 
helioseismology~\cite{Brown:2002ih} ($L_{1A} = 4.8\pm6.7$~fm$^3$).  
It should be noted that, in all of these cases, the stated uncertainty is 
larger than the measured value itself.  In contrast, the MuSun experiment 
proposes to measure $L_{1A}$ with an uncertainty of $\pm 1.25$~fm$^3$.

The pionless EFT calculation is in
principle only valid for low-energy processes, where the
momentum scale $Q << m_\pi$.
However, for muon capture, $Q = 102.1$~MeV~$\approx m_\pi$, so this theory
is not clearly appropriate.  Indeed, a component of the capture rate involves
the emission of neutrons at energies up to $Q/2 = 51$~MeV.  This part 
would not be described well by the pionless model, as it would necessarily 
include meson exchange currents.  Consequently, chiral perturbation
theory is also applied~\cite{Ando:2001es} to study muon capture on the 
deuteron.  In this theory, the LEC that describes the two-nucleon 
axial current is known as $\hat{d}^R$.   The MuSun data will also constrain
this parameter. 

MuSun will reuse a substantial amount of equipment and expertise from the MuCap
experiment; the basic disappearance method technique will remain essentially
unchanged, as will the electron detector system.
The most significant changes to the experiment arise from
the different kinetics of the muon-deuteron system relative to the
muon-proton system.  The deuteron has spin 1, so the allowed total-spin states 
for the muonic atom are the doublet ($J=1/2$) and the quartet ($J=3/2$).  
Initially, these
states are populated statistically, with 2/3 of the atoms in the quartet
state.  The desired measurement is from a pure atomic doublet state; 
unfortunately, the quartet-to-doublet transition rate $\lambda_{qd}$ is not 
sufficient to reach this state quickly (relative to the muon lifetime) at 
density $\phi=$1\%.  A higher density is needed to accelerate the spin-flip 
transition.  However, this increased density also increases the rate 
of $d-\mu-d$ molecular formation to an undesirable level; to compensate for 
this effect, we must reduce the temperature.  The optimal target conditions 
appear to be a density $\phi=5\%$ with a 
temperature $T=30$~K (rather than room temperature).  Consequently, a cryogenic
gas target system and a TPC that is able to operate under these
conditions are required.  A preliminary design for the chamber uses a pad 
plane approach; studies are in progress to determine its ideal 
geometry.

Neutron detector systems will be employed to observe both 
fusion ($d+d+\mu^- \rightarrow ^3$He$ + n + \mu^-$) and 
capture ($\mu^- + d \rightarrow n + n + \nu_\mu$) neutrons.
These two processes may be separated based on the neutron 
energy: all fusion electrons have a kinetic energy of 2.45~MeV, while some
capture neutrons reach as high as 53~MeV.   The fusion process is a 
sentinel of molecular formation; by monitoring these rates as a function
of time after the muon stops, the calculated atomic and molecular kinetics
can be verified.

A staged strategy has been formulated for the implementation of
the MuSun experiment.  In 2008, a prototype of the new TPC will be 
constructed and operated at room temperature in the muon beam at PSI,
with the engineering goal of demonstrating the performance of its pad 
geometry.  Important studies of systematic errors related to gas
impurities will also take place, along with a measurement of the 
residual polarization of the muon-deuteron atom.  By late 2009, the 
TPC should be ready for cryogenic operation, and it is anticipated that
the production data will be accumulated in 30~weeks of beam time over 
a two-year period.  A total of $1.8\times 10^{10}$ negative muon decays 
will be observed.  The positive muon lifetime will also be measured as a 
instrumental check in the new apparatus, requiring $1.2\times 10^{10}$ 
positive muon decays to give similar precision.

\begin{acknowledgement} 
The author would like to thank his colleagues in the MuCap and MuSun
collaborations for their support, as well as the organizers of the
2007 Advanced Studies Institute on Symmetries and Spin for an enjoyable
meeting.  Summer travel funds (for which the
author is particularly grateful) were provided by a National Science 
Foundation Research Opportunity Award supplement through the University 
of Illinois at Urbana-Champaign.
\end{acknowledgement}

% \bibliography{mucap_spin07}

\begin{thebibliography}{24}

\bibitem{muCapCollaboration}
{\nohyphens{MuCap Collaboration: V.A.~Andreev, B.~Besymjannykh, A.A.~Fetisov,
  V.A.~Ganzha, V.I.~Jatsoura, P.A.~Kravtsov, A.G.~Krivshich, M.~Levchenko,
  E.M.~Maev, O.E.~Maev, G.E.~Petrov, G.N.~Schapkin, G.G.~Semenchuk,
  M.A.~Soroka, V.~Trofimov, A.A.~Vasilyev, A.A.~Vorobyov,
  M.E.~Vznuzdaev~(Petersburg Nuclear Physics Institute, Gatchina, Russia);
  P.U.~Dick, A.~Dijksman, J.~Egger, D.~Fahrni, M.~Hildebrandt, A.~Hofer,
  L.~Meier, C.~Petitjean, R.~Schmidt~(Paul Scherrer Institut, Villigen,
  Switzerland); T.~Banks, T.A.~Case, K.M.~Crowe, S.J.~Freedman, F.E.~Gray,
  B.~Lauss~(University of California, Berkeley, CA, U.S.A.); D.B.~Chitwood,
  S.~Clayton, P.~Debevec, D.W.~Hertzog, P.~Kammel, B.~Kiburg, S.~Knaack,
  R.~McNabb, F.~Mulhauser, C.S.~\"Ozben, D.~Webber, P.~Winter~(University of
  Illinois at Urbana-Champaign, IL, U.S.A.); L.~Bonnet, J.~Deutsch,
  J.~Govaerts, D.~Michotte, R.~Prieels~(Universit\'e Catholique de Louvain,
  Belgium); R.M.~Carey, K.R.~Lynch~(Boston University, Boston, MA, U.S.A.);
  T.~Gorringe, V.~Tishchenko~(University of Kentucky, Lexington, KY, U.S.A.).}}

\bibitem{muSunCollaboration}
{\nohyphens{MuSun Collaboration: V.A.~Andreev, V.A.~Ganzha, P.A.~Kravtsov,
  A.G.~Krivshich, M.~Levchenko, E.M.~Maev, O.E.~Maev, G.E.~Petrov,
  G.N.~Schapkin, G.G.~Semenchuk, M.A.~Soroka, A.A.~Vasilyev, A.A.~Vorobyov,
  M.E.~Vznuzdaev~(Petersburg Nuclear Physics Institute, Gatchina, Russia);
  D.W.~Hertzog, P.~Kammel, B.~Kiburg, S.~Knaack, F.~Mulhauser,
  P.~Winter~(University of Illinois at Urbana-Champaign, IL, U.S.A.);
  M.~Hildebrandt, B.~Lauss, C.~Petitjean~(Paul Scherrer Institut, Villigen,
  Switzerland); T.~Gorringe, V.~Tishchenko~(University of Kentucky, Lexington,
  KY, U.S.A.); R.M.~Carey, K.R.~Lynch~(Boston University, Boston, MA, U.S.A.);
  R.~Prieels~(Universit\'e Catholique de Louvain, Belgium); F.E.~Gray~(Regis
  University, Denver, CO, U.S.A.); A.~Gardestig, K.~Kubodera,
  F.~Myhrer~(University of South Carolina, Columbia, SC, U.S.A.).}}

\bibitem{Bernard:1995dp}
V.~Bernard, N.~Kaiser, U.G. Meissner, Int. J. Mod. Phys. \textbf{E4}, 193
  (1995)

\bibitem{Gorringe:2002xx}
T.~Gorringe, H.W. Fearing, Rev. Mod. Phys. \textbf{76}, 31 (2004)

\bibitem{AlberigiQuaranta:1969ip}
A.~Alberigi~Quaranta et~al., Phys. Rev. \textbf{177}, 2118 (1969)

\bibitem{Bystritsky:1974}
V.M. Bystritsky et~al., Zh. Eksp. Teor. Fiz. \textbf{66}, 43 (1974), {Sov.}
  Phys. JETP {\bf 39}, 19 (1974)

\bibitem{Bardin:1980mi}
G.~Bardin et~al., Nucl. Phys. \textbf{A352}, 365 (1981)

\bibitem{Wright:1998gi}
D.H. Wright et~al., Phys. Rev. \textbf{C57}, 373 (1998)

\bibitem{Bardin:1981cq}
G.~Bardin et~al., Phys. Lett. \textbf{B104}, 320 (1981)

\bibitem{Clark:2005as}
J.H.D. Clark et~al., Phys. Rev. Lett. \textbf{96}, 073401 (2006)

\bibitem{Bakalov:1980fm}
D.D. Bakalov, M.P. Faifman, L.I. Ponomarev, S.I. Vinitsky, Nucl. Phys.
  \textbf{A384}, 302 (1982)

\bibitem{Andreev:2007wg}
V.A. Andreev et~al. (MuCap Collaboration), Phys. Rev. Lett. \textbf{99}, 032002
  (2007)

\bibitem{Chitwood:2007pa}
D.B. Chitwood et~al. (MuLan Collaboration), Phys. Rev. Lett. \textbf{99},
  032001 (2007)

\bibitem{Czarnecki:2007th}
A.~Czarnecki, W.J. Marciano, A.~Sirlin, Phys. Rev. Lett. \textbf{99}, 032003
  (2007)

\bibitem{Ganzha:2007uk}
V.A. Ganzha et~al., Nucl. Instrum. Meth. \textbf{A578}, 485 (2007)

\bibitem{2007NIMPB.259....7S}
H.A. {Synal}, M.~{Stocker}, M.~{Suter}, {Nucl. Instr. Meth.} \textbf{B259}, 7
  (2007)

\bibitem{2004ITPS...32.1932B}
M.J. {Barnes}, G.D. {Wait}, IEEE Transactions on Plasma Science \textbf{32},
  1932 (2004)

\bibitem{Aharmim:2007nv}
B.~Aharmim et~al. (Sudbury Neutrino Observatory Collaboration), Phys. Rev.
  \textbf{C75}, 045502 (2007)

\bibitem{muSunProposal}
V.A. Andreev et~al. (MuSun Collaboration), \emph{Muon capture on the deuteron:
  the {MuSun} experiment},
  {http://www.npl.uiuc.edu/exp/musun/documents/prop07.pdf}

\bibitem{Schiavilla:1998je}
R.~Schiavilla et~al., Phys. Rev. \textbf{C58}, 1263 (1998)

\bibitem{Chen:2002pv}
J.W. Chen, K.M. Heeger, R.G.H. Robertson, Phys. Rev. \textbf{C67}, 025801
  (2003)

\bibitem{Butler:2002cw}
M.~Butler, J.W. Chen, P.~Vogel, Phys. Lett. \textbf{B549}, 26 (2002)

\bibitem{Brown:2002ih}
K.I.T. Brown, M.N. Butler, D.B. Guenther (2002), \texttt{nucl-th/0207008}

\bibitem{Ando:2001es}
S.~Ando, T.S. Park, K.~Kubodera, F.~Myhrer, Phys. Lett. \textbf{B533}, 25
  (2002)

\end{thebibliography}

\end{document}